**Insights into carbon nanotube nucleation: Cap formation governed by catalyst interfacial step flow**

*Rahul Rao[1], Renu Sharma[2], Frank Abild-Pedersen[3], Jens K. Nørskov[3,4], Avetik R. Harutyunyan[1*]*

[1]Honda Research Institute USA Inc., Columbus, Ohio, 43212, USA

[2]Center for Nanoscale Science and Technology, National Institute of Standards and Technology, Gaithesburg, Maryland, 20899, USA

[3] SUNCAT Center for Interface Science and Catalysis, SLAC National Accelerator Laboratory, Menlo Park, CA, USA

[4]Department of Chemical Engineering, Stanford University, Stanford, CA, 94305, USA

**In order to accommodate an increasing demand for carbon nanotubes (CNTs) with desirable characteristics one has to understand the origin of helicity of their structures. Here, through in situ microscopy we demonstrate that the nucleation of a carbon nanotube is initiated by the formation of the carbon cap. Nucleation begins with the formation of a graphene embryo that is bound between opposite step-edges on the nickel catalyst surface. The embryo grows larger as the step-edges migrate along the surface, leading to the formation of a curved carbon cap when the steps flow across the edges of adjacent facets. Further motion of the steps away from the catalyst tip with attached rims of the carbon cap generates the wall of the nanotube. Density Functional Theory calculations bring further insight into the process, showing that step flow occurs by surface self diffusion of the nickel atoms via a step-**

---

[*] Corresponding author – A.R. Hatutyunyan, Email: aharutyunyan@honda-ri.com

**edge attachment-detachment mechanism. Since the fact that cap forms first in the sequence of stages involved in nanotube growth, we suggest that it originates the helicity of the nanotube. Therefore, the angular distribution of catalyst facets could be exploited as a new parameter for controlling the curvature of the cap and, presumably, the helicity of the nanotube.**

**Introduction**

Controlled growth of carbon nanotubes with desired properties is imperative for their unique applications,[1-3] which requires a full understanding of their atomistic growth mechanism. Towards this end, earlier work by Helveg *et al.*[4] was a substantial contribution to the field where the early stage of catalytic formation of graphitic carbon layers was captured in situ by environmental transmission electron microscopy (ETEM) and provided important hints into the growth mechanism of carbon nanofibers. Although the growth mechanism of a carbon nanofiber cannot be easily extended to crystalline nanotubes,[5] these studies revealed reaction-induced reshaping of Ni catalyst particles by the restructuring of monoatomic step edges. The experimental observations, combined with theoretical modeling, suggested that step-edge sites act as the preferential growth centers for graphitic layers on the Ni catalyst surface[4,6,7].

The nucleation of a carbon nanotube within an ETEM was observed later by the injection of carbon atoms from graphitic shells surrounding the metal catalyst particle into the body of the particle by electron beam irradiation (Ref. 8). This event was also accompanied by dynamic morphological changes of catalyst particle (Fe) during tube growth, suggesting wetting-driven deformation of the particle tip into a convex dome as a necessity for the formation of the carbon nanotube cap[8]. In parallel, other ETEM studies have revealed more insights into the CNT growth mechanism with observations of catalyst reconstruction and surface steps bounded by nanotube rims[9-14]. The majority of the researchers that have targeted this problem attest to the key roles of catalyst size and structure in the formation of nanotube symmetry based on observations such as the correlation between the nanotube wall basal plane and the structure of the corresponding facet on the catalyst, the impact of catalyst composition, phase and pretreatment conditions on the structure of the nanotubes, and the selectivity of the growth kinetics for nanotubes with various chiral indices.[14-23]

In spite of the large number of in situ studies capturing the early stages of CNT growth, specifically, the nucleation of the carbon nanotube cap has never been observed, and its mechanism still remains unclear. The structure of the cap is inherently different from a bent graphene layer and its uniqueness implies that a given cap structure fits only one particular nanotube, whereas a given nanotube wall can have thousands of caps[24,25]. In other words the carbon cap symmetry can dictate the nanotube helicity depending on when and how it forms during the initial stages of nanotube nucleation. Thus the real-time observation of cap formation and revealing of its mechanism are essential for understanding the origin of the helicity in the nanotube.

Two of the experimental challenges that have thus far precluded this are the difficulty in locating a particular catalyst that is capable of CNT nucleation, and the very short nucleation times, especially in the case of single-walled nanotubes. In this work we overcame these obstacles by imaging of cap nucleation of the innermost tube during catalytic growth of multi-walled nanotube (MWNT) at a low temperature (520 ˚C). Precise knowledge of the catalyst particle tip location made it possible to detect the initial stage of growth of the innermost tube in the MWNT. The space between the surface of the catalyst particle and the previously grown nanotube was thus used as a nanoscale reactor inside which we observed the nucleation of the new inner tube. We found that the formation of the nanotube cap precedes that of the wall in the sequence of steps occurring during the nanotube nucleation process. Moreover, we observed that the cap is formed by the evolution of a graphene embryo, which is constrained between opposite steps on the catalyst particle surface. The steps flow through adjacent facets on the catalyst tip, introducing curvature on the graphene embryo and leading to cap formation. Subsequent motion of the steps towards the open end of the particle causes the elongation of the nanotube wall. The sequence of these processes suggests that the carbon cap structure dictates the symmetry or helicity of the nanotube. Furthermore, the revealed growth mechanism implies that the helicity could be controlled by variation of carbon cap curvature via the exploitation of interfacial angles between the corresponding adjacent facets on catalyst nanoparticles.

RESULTS

**Observation of carbon nanotube cap nucleation**

The MWNTs were grown on Au-doped Ni nanoparticles (10 to 15 nm) at 520 $^o$C using acetylene ($C_2H_2$) as the carbon source (see Methods for experimental details and Ref. 26 for details on the effect of Au doping). Under these conditions, CNT growth occurred by the tip-growth mode, eliminating the possibility of catalyst shape reconstruction induced by interaction with a substrate. Furthermore, the low growth temperature (520 °C) provided a relatively slow growth rate (≈1 nm/s) and allowed the catalyst particle to preserve its crystallinity during CNT growth. Hence, the catalyst shape changes observed during the experiment were induced only by the decomposition of the carbon feedstock and adsorption of carbon on the catalyst surface, not due to substrate effects or liquefaction of the catalyst. Figs. 1a and 1b show snapshots captured just after cap lift-off and during elongation of the innermost tube (indicated by white arrows) within the MWNT. The particle shape changes accompanying tube growth can be clearly seen. Fig. 1c shows a high resolution TEM image of the crystalline catalyst particle, with the spacing between the lattice fringes (0.25 nm) corresponding to crystalline $Ni_3C$. Furthermore, fast Fourier transform (FFT or diffractogram) analysis of the crystalline particle (inset in Fig. 1c) confirms its structure to be $Ni_3C$. The facets on the particle are assigned as $(\bar{1}2\bar{1})$, $(\bar{1}0\bar{3})$, and $(1\bar{1}0)$ based on measured and calculated interfacial angles for $Ni_3C$ (Fig. 1d), and are listed in Table 1.

Time-resolved images from a digital video sequence capturing the nucleation of the inner tube within a MWNT (Supplementary Information, Video 1) are shown in Fig. 2. A high video frame rate (15 $s^{-1}$), coupled with the low CNT growth rate provided the temporal resolution to observe the nucleation of the innermost tube and associated catalyst morphology changes. We chose the time $t = 0$ as just before the graphene embryo formation was observed on the surface of the catalyst particle. As shown by the sequence of images and the corresponding schematics in Fig. 2, the cap formation of the new inner nanotube begins with the formation of a graphene embryo on the ($\bar{1}21$) facet of the catalyst particle (Fig. 1a). The embryo is bound on both sides by steps on the surface of the catalyst particle (indicated by the white arrow in Fig. 2). The two steps have opposite signs and start to flow in opposite directions on the particle surface, causing elongation of the graphene embryo. Such motion of graphene layers bounded by catalyst steps is also reported in Ref. 4. Remarkably, by $t = 0.3$ s (Fig. 2b), the step on the right reaches the end of the facet and crosses over to the adjacent facet which is at an angle of ≈60° to the ($\bar{1}\bar{2}1$) facet. The graphene embryo can be seen clearly attached to the step (indicated by the black arrows in Fig. 2). The interfacial motion of the step across the particle tip surface introduces curvature into the growing graphene embryo. This curvature increases as the step crosses over another adjacent facet (Fig. 2f) leading to nanotube cap formation. Over the next few seconds both steps keep moving simultaneously away from particle tip, leaving behind the nanotube cap bound to the particle. Detachment (lift-off) of the cap occurs after several seconds due to reconstruction of the particle facets under the cap. The nascent nanotube can be seen in Fig. 2g. The video used for analysis and described below is typical for the videos we have described previously[13,26]. However here for the first time we present the nucleation of a nanotube cap. We were also able to observe nucleation events of different inner tubes at later times during the recording of the video albeit with lower resolution due to increasing vibration of the catalyst particle as the overall length of the nanotube increased over time.

**Step flow**

The observation of nanotube cap formation described above highlights the importance of the angular inter-relationships between adjacent facets on a catalyst surface, which introduce curvature in the evolving graphene embryo and influence the structure of the nanotube cap. This view is a significant advance over the current status quo, where typically only one facet is considered for structure control via planar epitaxy. Hence our observations imply that one needs to consider cap formation and symmetry control over a 3D particle, rather than the present simplified 2D approach.

We first consider the process of carbon adsorption on the catalyst particle at growth temperature. It is established that carbon adsorption and its degree of coverage lead to surface reconstruction of the catalyst particle, which varies depending on the structure of the facets[13,27]. Since the arrangement of steps on the surface defines the particle morphology, step flow is one of the most likely mechanisms through which the surface reconstruction takes place. The presences of steps on crystal surfaces are common and can occur by thermal fluctuations of edge atoms, leading to their detachment[28]. Such steps and kinks are also likely to be present on the surfaces of smaller particles[29]. Furthermore, surface steps are known to be high reactivity sites[7,30] and can induce adatom (e.g. carbon) adsorption[28], since the carbon binding energy to the Ni step is larger than the energy cost for step formation. Hence, in our experiment carbon adsorption, either on steps or on terraces, followed by graphene embryo formation causes a variation in the surface energies of facets. Consequently, the catalyst surface tries to equilibrate via reconstruction through the step flow, causing the development of the attached graphene embryo into a nanotube cap (Fig. 2).

As mentioned above, one of the distinguishing features of CNT nucleation from the formation of graphene layers during growth of carbon fibers is that in order to produce a nanotube the simple extension or bending of the graphene embryo around the catalyst particle is not enough – It needs to form a cap on the catalyst tip. As shown in Fig. 2, cap formation is realized only when the steps (with the attached graphene embryo) flow over adjacent facets on the catalyst tip. This unique process puts certain restrictions on the symmetries of adjacent facets and thus on the corresponding surface orientations in order to maintain the growing nanotube rather than encapsulation of the catalyst particle by a graphitic carbon layer. The interfacial angular distribution between catalyst particles is a convenient parameter that can be used in a representation of both particle surface curvature and the symmetry for nucleation of the carbon cap. For instance, for a Ni particle (FCC structure), the (100), (111) and (110) vertices of the stereographic triangle are achiral surfaces [consisting of stepped and close-packed surfaces: with Miller indices (hkk), (hhl) or (hk0)], while points within the triangle represent chiral surfaces (consisting of stepped, kinked and close-packed surfaces, with Miller indices h>k>l≠0). However, the final symmetry of the cap also depends on the interplay between the energies that are needed for the incorporation of pentagons (isolated or adjacent pairs) with various configurations on the growing graphene embryo[25,31].

Density Functional Theory (DFT) calculations of graphene growth on metal surfaces have shown that the preferential nucleation sites (step edges, kinks or terraces) for a graphene embryo depend on the corresponding facet symmetries on the catalyst surface[32]. In addition, the lowest critical size of a graphene embryo also depends on the facet (step/terrace) symmetry, which was concluded based on competition between the energy cost of graphene embryo edges and formation of thermodynamically stable bulk graphene layer[7,32,33]. Hence, from a thermodynamic viewpoint, cap formation can occur on a surface of the particle tip via step flow if the symmetries of adjacent facets and their step/kink edges satisfy the conditions where carbon atoms bind most favorably to the step/kink edges. In addition, the initial step-bound graphene embryo must be stable so that it can grow. Finally, the interrelated sequence of facet symmetries should be favorable for cap lift-off upon reconstruction instead of formation of layered carbon[13].

Next, we would like to point out interesting events that were observed during the growth of the MWNT in the ETEM. In addition to step-mediated initial nucleation of a nanotube cap, we observe three different morphologies formed on the catalyst surface as a result of step flow processes resulting in termination of nanotube growth. Fig. 3a demonstrates the first scenario where the step flow (with the attached nanotube wall) towards the carbon-free end of the particle caused the eventual flattening of the step. This in turn leads to the detachment of the nanotube wall (black arrow), and consequently, termination of growth. This type of growth termination is particularly more prevalent in bamboo-shaped or herringbone-shaped nanotubes and was observed also in case of carbon fiber growth.[4] The second scenario we observe is that nanotube rim remains attached even after flattening of corresponding step, causing the attached wall to become bent (Fig.3b), which has also been observed previously in MWNTs[6,8]. The many walls of a MWNT are also most commonly observed attached to several densified steps at the end of growth (Fig.3c, d). In this case the terrace lengths are significantly shorter in comparison to their lengths at the beginning stages of CNT growth. We attribute this observation to the phenomenon known as step bunching, which occurs when the steps on the vicinal facets of a crystal surface become perturbed due to kinetic instabilities that destabilize a uniform step train[34-36], causing the steps to bend or aggregate together. In our case one could consider destabilization of the step train as a result of adsorbed carbon adatoms and thereby model the step bunching phenomena during CNT growth by applying impurity-induced step bunching mechanisms proposed first by Frank[34] and further developed by others later[37-39]. However, these models assume non-interacting impurities in front of a step that impedes its motion. In the case of CNT growth, the carbon adatoms (impurities) bond with each other and form a graphene layer that is attached to the steps and covers the entire upper and lower terraces depending on the instant of growth. Thus the models described above cannot be applied. Theoretical modeling of our observations would surely provide greater understanding of the cap nucleation mechanism, yet this would be a daunting task for the present work considering the complications and the fact that it took decades for the development of existing various models.

DISCUSSION

Although a complete theoretical modeling of graphene-bound step flow kinetics is not possible at this time, it is still possible to discuss the path of step flow during Ni surface reconstruction. As shown in Figs. 2 and 3, a graphene layer (cap or wall) is bound to steps and covers the terrace. Since the binding energy between Ni and carbon based on experimental results (≈7 eV)[40] appears stronger than the Ni-Ni bond (≈2.3 eV)[41], step motion could occur by surface self-diffusion or attachment-detachment of Ni atoms from the step only during the growth of a MWNT. There are two ways in which the step-based growth of MWNT can proceed and both involve the diffusion of Ni atoms away from the step-edge. Fig. 4 shows the schematics of the different possible processes: (1) bulk diffusion (Fig. 4a), and (2), diffusion of a Ni atom under the graphene sheet (blue arrow in Fig. 4b) or up on the terrace under another sheet (red arrow in Fig. 4b). We have performed DFT calculations of Ni atom diffusion on Ni (111) surfaces to identify which of the two possible routes are most feasible. We note that while the particle studied here is in the carbide form, we considered pure metal catalyst for the calculation since the necessary parameters are readily available. Binding energies of Ni on a $Ni_3C$ surface are similar to the binding energies of Ni on clean Ni(111) and hence the effect on the surface diffusion on Ni will be limited. The large energy difference between having Ni on the surface and in the bulk is maintained and would therefore suggest a kinetic hindrance for this process even with the simplified model. Furthermore, we rely on the fact that similar morphological changes of the particle have been observed with both carbide and metallic catalyst composition[11,13,27,42,43]. For the metal particle-mediated process we have found that diffusion of a Ni atom between two bulk, or subsurface interstitial sites is associated with a barrier of less than 0.2 eV in both processes. However, the energy differences between having a Ni atom on the Ni (111) surface or in the subsurface or bulk interstitial sites are 2.54 eV and 4.8 eV, respectively. This indicates that surface diffusion of Ni atoms is more likely to play a significant role in the catalyzed growth of carbon nanotubes. In a prior DFT study it was shown that the graphene sheet enhanced the stability of atomic Ni on the surface and that the attachment-detachment of Ni from the step-edge during carbon incorporation in the growing fiber had a barrier of less than 0.6 eV[4]. Hence, the surface mediated process is associated with much lower energy barriers and the mechanisms shown in Fig. 4b are enough to facilitate continued growth of the carbon nanotubes.

**Conclusion**

Based on our results, we describe the CNT nucleation process by the following sequence: 1) Formation of steps on the catalyst surface via carbon adsorption (or precipitation of carbon at pre-existing steps); 2) Growth of a graphene embryo constrained by steps with opposite signs; 3) Introduction of curvature into the graphene embryo and formation of the nanotube cap by step flow over adjacent facets on the catalyst tip; 4) Elongation of the cap and growth of the nanotube by step flow away from the catalyst tip.

We draw the following conclusions from this model: 1) The angular interrelationships between adjacent facets on catalyst tip define the feasibility of cap nucleation; 2) Cap structure governs the symmetry of nanotube and thereby originates its helicity 3) Step flow during MWNT growth occurs via surface self-diffusion of catalyst atoms; 4) In general MWNT growth is accompanied by step bunching phenomenon on catalyst surface.

A word of caution is needed however: depending on the catalyst composition and symmetry of particular facets, there could also be a scenario where carbon binding with the terrace is more preferable than with the step edge[32]. Theoretically this may lead to a different sequence of processes during the initial growth stage of a nanotube, and probably a different mechanism for helicity formation. Nevertheless, our actual observations imply a self-consistent relationship between catalyst reconstruction and cap nucleation in the manner that the interrelationships between facet structures define the feasibility of cap formation and its curvature, while the nucleated cap is responsible for the structural symmetry of the nanotube. Hence, our results suggest control over the interfacial angles between facets on the catalyst surface as the path that can presumably lead to helicity-controlled growth of CNTs. The viability of this path depends on our capability of synthesizing stable catalyst particles with a fine degree of control over its diameter and interfacial angular distribution.

**Methods**

Thin films of Ni (≈1 to 2 nm thick, with a small amount of Au) were first deposited on perforated $SiO_2$ films supported on 200 mesh Mo TEM grids by physical vapor deposition[26]. The grids were loaded on a TEM heating holder and introduced to the ETEM column. Upon heating (> 200 °C) the films dewetted from the $SiO_2$ substrate to form 4 to 7 nm diameter particles. The size of the particles did not change appreciably upon further heating to the reaction temperatures used (520 °C). Samples were held at the reaction temperature for ≈25 min. in order to stabilize the temperature and fully reduce any NiO (if present) to Ni. $C_2H_2$ was then introduced into the ETEM to induce CNT growth. A pressure of ≈0.4 Pa was maintained during the growth period of 15 min.

After every in situ growth experiment, we recorded images from regions not exposed to the electron beam during the experiment at room temperature and in high vacuum. We compared the images of the tubes/catalyst particle with the ones recorded during growth and find no difference. Therefore, based on this level general capability and fact that our irradiation was carried out at beam current densities ~10A/cm$^2$, while even in case of 10$^3$-10$^5$A/cm$^2$ current densities atomic displacement was observed only from MWNTs but not from metal particles due to the high displacement threshold energy in metals[44-46] we can exclude electron irradiation effects during the nanotube growth process under our experimental conditions.

All calculations were performed in the planewave DFT code Quantum Espresso[47]. Ultra-soft pseudopotentials were used for carbon and nickel[48]. For the exchange and correlation the semiempirical BEEF-vdW functional have been used which specifically includes van der Waals dispersion interactions[49]. A 4x4x6 super cell model was used for the calculations of subsurface diffusion of interstitial Ni were the top most three layers were allowed to relax fully. A 4x4x4 super cell model was used for the calculation of bulk diffusion of Ni and all atoms in the cell were allowed to relax. The Brillouin zones were sampled with 4x4x1 and 4x4x4 Monkhorst-Pack *k*-points[50], respectively. The kinetic energy cutoff for the plane wave basis sets was 500 eV, which have been chosen to ensure convergence within 0.1eV. To find the transition state for each diffusion reaction we used the Nudged-Elastic-Band method.

## AUTHOR CONTRIBUTIONS

R.R. analyzed the data, R.S. performed the TEM experiments and helped with data analysis, F. A-P. and J.K.N. performed DFT calculations, A.R.H. initiated and designed the research and analyzed the data. R.R. and A.R.H. wrote the paper.

## ACKNOWLEDGMENTS

We thank A. Zangwill for helpful discussions and advice on this manuscript. F.A-P and J.K.N. acknowledges financial support from the U.S. Department of Energy, Office of Basic Energy Sciences to the SUNCAT Center for Interface Science and Catalysis. This research was supported by the Honda Research institute USA Inc.

## ADDITIONAL INFORMATION

The authors declare no competing financial interests.

## FIGURE LEGENDS

**Figure 1. *In situ* TEM images recorded during MWNT growth.** (a,b) High-resolution TEM images recorded just after lift-off of the innermost tube inside a MWNT (a), and during elongation of the tube (b). The innermost tube is indicated by a white arrow in (a) and (b). (c) High magnification view of the catalyst particle tip just after lift-off of the innermost tube inside the MWNT. The FFT (inset in c) from the particle can be indexed to the $(1\bar{1}0)$ spacing from $Ni_3C$. (d) The same view as in (c) with the facet structures indicated. The structures of the neighboring facets were estimated from the angles between the planes according to the $Ni_3C$ crystal structure. All scale bars in the figure are 2 nm.

**Figure 2. Image sequence captured from Video 1 showing nanotube cap formation.** Images (a-g) show the process of nanotube cap formation followed by lift-off. Schematics are included with each figure to show the elongation of the graphene embryo bound to steps on the $(\bar{1}\bar{2}1)$ facet of the catalyst particle. The white and black arrows indicate the step and nanotube cap, respectively. The scale bar is 5 nm.

**Figure 3. Step flow-induced termination of CNT growth and step bunching.** (a) TEM image showing detachment of the outer wall (indicated by the black arrow) and termination of growth due to flattening of the step. (b) TEM image showing bending of the nanotube wall at the attachment point to the step. The scale bar is 5 nm. (c) High-resolution image captured the agglomeration of the steps with attached MWNT walls (d) The same view as in (c) with a dotted line outlining the step structure as a guide to the eye. The structure of the step closest to the particle surface (indicated by the white arrow) can be indexed as $(11\bar{3})$. Scale bar in the figure is 2 nm.

**Figure 4. Nickel atom diffusion during CNT growth.** 3D Schematics showing a Ni(111) surface with monoatomic (100) steps facilitating the growth of carbon nanofibers. We have explicitly shown the pathways for, (a) subsurface (or bulk) diffusion and, (b) surface diffusion of Ni atoms during step flow growth when the step is connected to a MWNT. The latter involves two possible scenarios in which both are identical in their final state energy. The Ni step-edge atom can be pushed onto the upper terrace (red arrow) or under the growing graphene layer (blue arrow). CNT growth is expected to proceed via the process involving Ni detachment under the graphene layer due to the enhanced stability induced by the graphene layer.

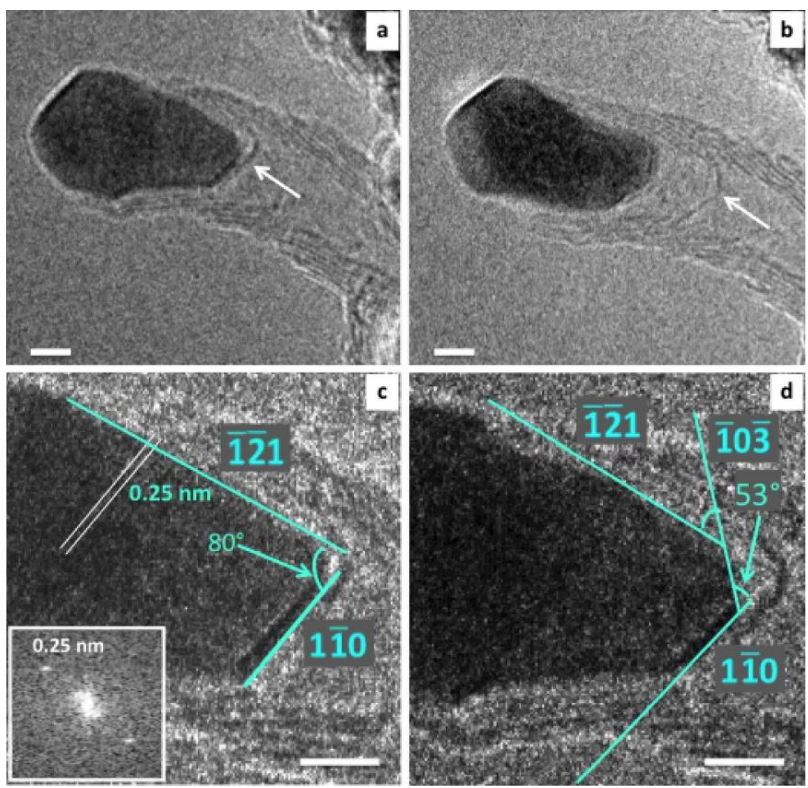

**Figure 1.** *In situ* **TEM images recorded during MWNT growth.** (a,b) High-resolution TEM images recorded just after lift-off of the innermost tube inside a MWNT (a), and during elongation of the tube (b). The innermost tube is indicated by a white arrow in (a) and (b). (c) High magnification view of the catalyst particle tip just after lift-off of the innermost tube inside the MWNT. The FFT (inset in c) from the particle can be indexed to the (1$\bar{1}$0) spacing from Ni$_3$C. (d) The same view as in (c) with the facet structures indicated. The structures of the neighboring facets were estimated from the angles between the planes according to the Ni$_3$C crystal structure. All scale bars in the figure are 2 nm.

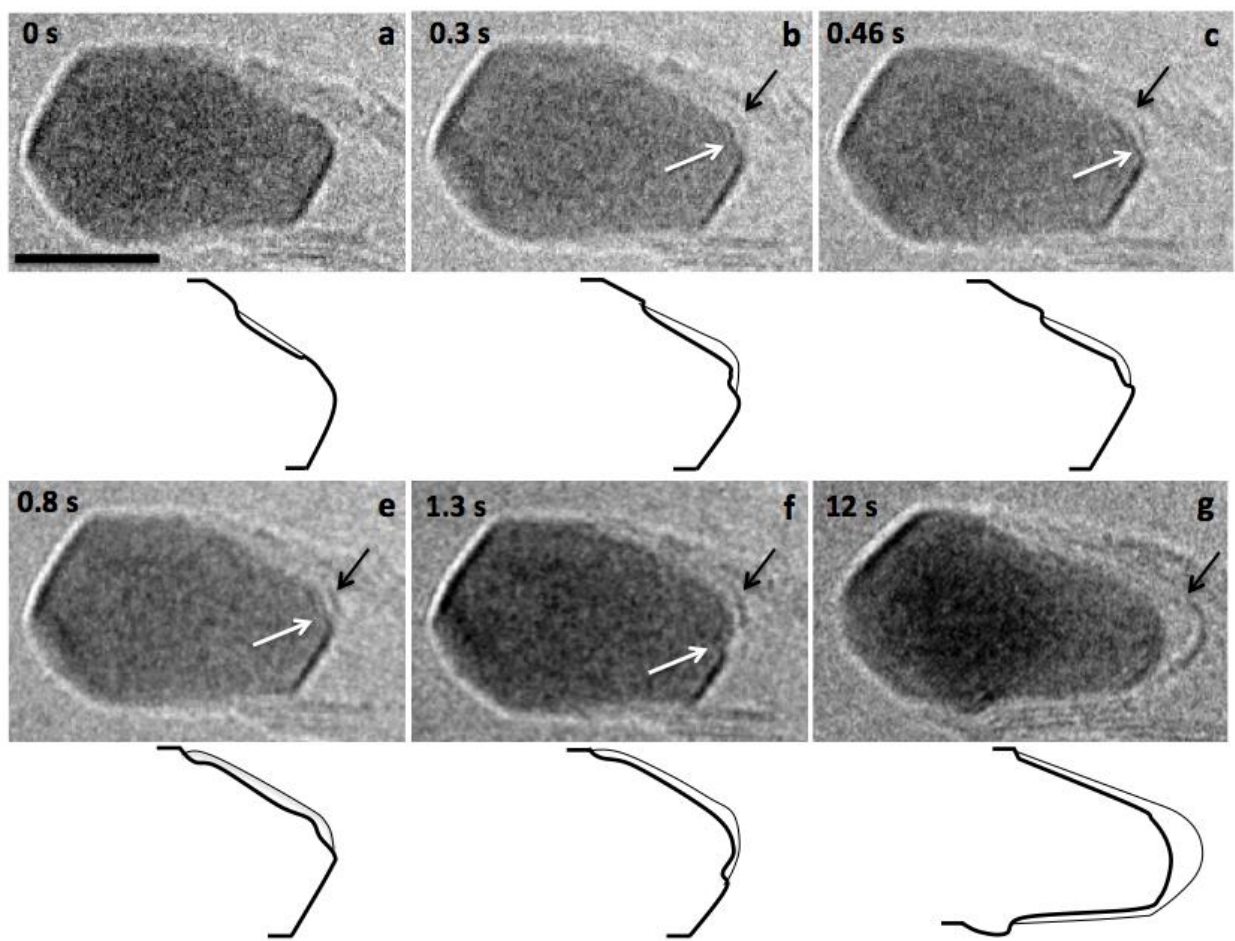

**Figure 2. Image sequence captured from Video 1 showing nanotube cap formation.** Images (a-g) show the process of nanotube cap formation followed by lift-off. Schematics are included with each figure to show the elongation of the graphene embryo bound to steps on the $(\bar{1}\bar{2}1)$ facet of the catalyst particle. The white and black arrows indicate the step and nanotube cap, respectively. The scale bar is 5 nm.

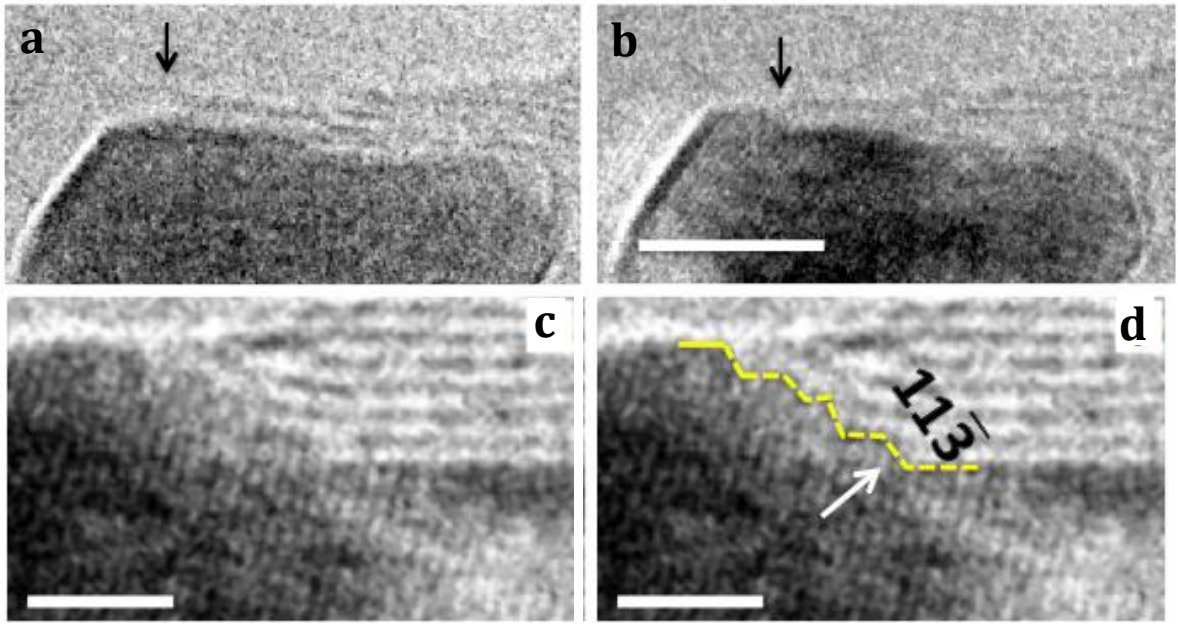

**Figure 3. Step flow-induced termination of CNT growth and step bunching.** (a) TEM image showing detachment of the outer wall (indicated by the black arrow) and termination of growth due to flattening of the step. (b) TEM image showing bending of the nanotube wall at the attachment point to the step. The scale bar is 5 nm. (c) High-resolution image captured the agglomeration of the steps with attached MWNT walls (d) The same view as in (c) with a dotted line outlining the step structure as a guide to the eye. The structure of the step closest to the particle surface (indicated by the white arrow) can be indexed as $(11\bar{3})$. Scale bar in the figure is 2 nm.

a 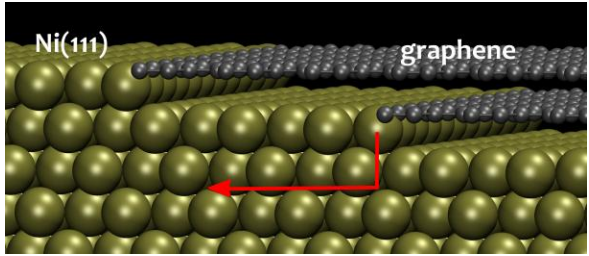 b 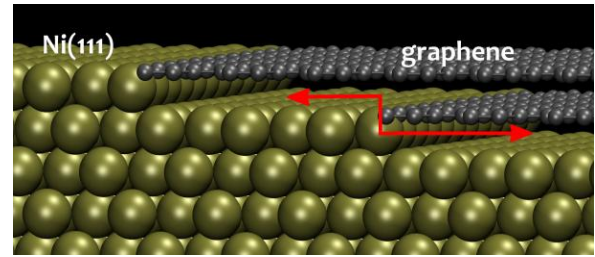

**Figure 4. Nickel atom diffusion during CNT growth.** 3D Schematics showing a Ni(111) surface with monoatomic (100) steps facilitating the growth of carbon nanofibers. We have explicitly shown the pathways for, (a) subsurface (or bulk) diffusion and, (b) surface diffusion of Ni atoms during step flow growth when the step is connected to a MWNT. The latter involves two possible scenarios in which both are identical in their final state energy. The Ni step-edge atom can be pushed onto the upper terrace (red arrow) or under the growing graphene layer (blue arrow). CNT growth is expected to proceed via the process involving Ni detachment under the graphene layer due to the enhanced stability induced by the graphene layer.

Table 1. Measured and calculated angles, and assigned facets for the particle shown in Fig. 1d.

| Measured Angles (°) | Calculated Angles (°) | Assigned facets |
|---|---|---|
| 53 | **51.9** | $(1\bar{1}0)/(\bar{1}0\bar{3})$ |
| 80 | **79.2** | $(1\bar{1}0)/(\bar{1}\bar{2}1)$ |
| 60 | **63.2** | $(\bar{1}0\bar{3})/(\bar{1}\bar{2}1)$ |